\documentclass[letterpaper]{article}
\usepackage{aaai19}  
\usepackage{times}  
\usepackage{helvet}  
\usepackage{courier}  
\usepackage[hyphens]{url} 
\usepackage{graphicx}  
\usepackage{placeins}

\usepackage{colortbl}
\usepackage{nth}
\usepackage{capt-of}
\usepackage{comment}
\usepackage{multirow}
\usepackage{floatrow}
\usepackage{float}
\usepackage{booktabs}
\usepackage{enumitem}
\usepackage{multicol}
\usepackage{graphicx}
\usepackage{subcaption}
\usepackage{amsmath}

\usepackage{arydshln}

\newcommand\blfootnote[1]{%
  \begingroup
  \renewcommand\thefootnote{}\footnote{#1}%
  \addtocounter{footnote}{-1}%
  \endgroup
}

\begin{document}
\title{NELA-GT-2019: A Large Multi-Labelled News Dataset for The Study
of Misinformation in News Articles}


\author{Maur\'{i}cio Gruppi, Benjamin D. Horne, and Sibel Adal{\i}\\
 Rensselaer Polytechnic Institute\\
 \{gouvem, horneb,adalis\}@rpi.edu
}

\maketitle

\begin{abstract} 
In this paper, we present an updated version of the \texttt{NELA-GT-2018} dataset~\cite{norregaard2019nela}, entitled \texttt{NELA-GT-2019}. \texttt{NELA-GT-2019} contains 1.12M news articles from 260 sources collected between January 1st 2019 and December 31st 2019. Just as with \texttt{NELA-GT-2018}, these sources come from a wide range of mainstream news sources and alternative news sources. Included with the dataset are source-level ground truth labels from 7 different assessment sites covering multiple dimensions of veracity. The \texttt{NELA-GT-2019} dataset can be found at: \url{https://doi.org/10.7910/DVN/O7FWPO}
\end{abstract}

\section{Introduction}\blfootnote{Data at: \url{https://doi.org/10.7910/DVN/O7FWPO}} 
A continued barrier to news veracity research is the availability of labeled news datasets. To sufficiently answer many research questions, news datasets must be both large in number of data points and timely. For example, machine learning studies not only require large, labeled data to train models, but also data that extends over long stretches of time to ensure models are accurate under concept drift. Other types of studies, such as mixed-method studies to understand disinformation tactics, news narratives, and the like, require data that is timely to ensure conclusions are adequately reached. Lastly, news veracity studies, both in machine learning and in computational social science, need broadly labeled data. News can misinform through methods other
than explicitly fabricated claims, hence having labels that are not only based on fact-checking, but also based on bias, consumer trust, and source behavior are needed. The dataset presented in this paper attempts to meet these goals. 

There have been multiple labeled news article datasets released recently, including the NELA-GT-2018 dataset \cite{norregaard2019nela}, the FA-KES dataset \cite{salem2019fa}, and the Golbeck et al. dataset  \cite{golbeck2018fake}. Other datasets have focused on social media data rather than news article data, including the FakeNewsNet dataset \cite{shu2018fakenewsnet} and the LIAR dataset \cite{wang2017liar}. In addition, several studies have released smaller, study specific datasets. 

While data curation has been an increased focus of researchers and journalist as of late, data must continue to be collected and labeled in order for timely research to occur. Hence, in this paper we present \texttt{NELA-GT-2019}, an update to the \texttt{NELA-GT-2018} dataset. Specifically, we continued our collection of the 194 news sources in the \texttt{NELA-GT-2018} dataset, as well as added 66 more sources. In total, \texttt{NELA-GT-2019} contains \textbf{260 news sources} with \textbf{1.12M news articles} published between January 1st, 2019 and December 31st, 2019. Additionally, we continued our collection of source-level labels from multiple news veracity assessment sites, including Media Bias/Fact Check (MBFC), Allsides, and PolitiFact.

In this short paper, we describe the key differences between the 2018 version and the 2019 version of the dataset. We also describe in detail the data collection method, ground truth collection method, and publicly available data formats. Lastly, we provide metadata and a discussion of use cases. 

\begin{figure*}[ht!]

    \begin{subfigure}{\textwidth}
        \centering
        \includegraphics[width=0.8\textwidth]{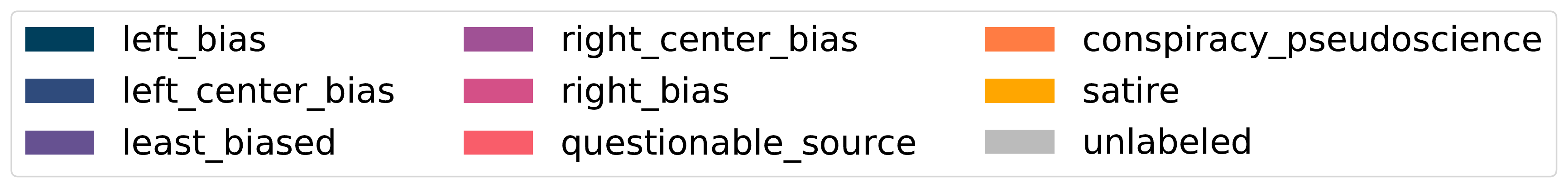}
    \end{subfigure}

    \begin{subfigure}{0.48\textwidth}
        \centering
        \includegraphics[width=\textwidth]{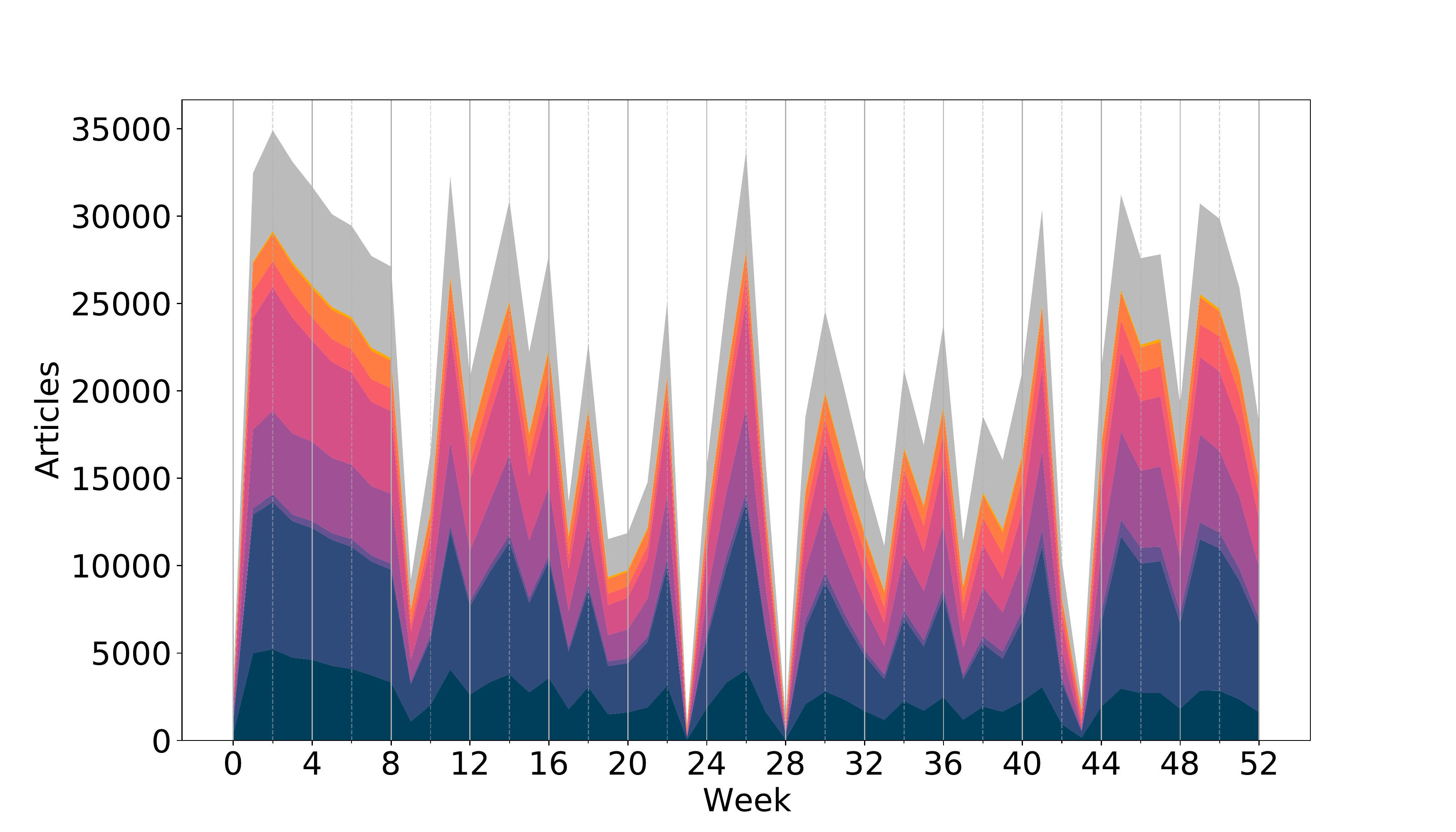}
        \caption{MBFC Categories}
    \end{subfigure}
    \begin{subfigure}{0.48\textwidth}
        \centering
        \includegraphics[width=\textwidth]{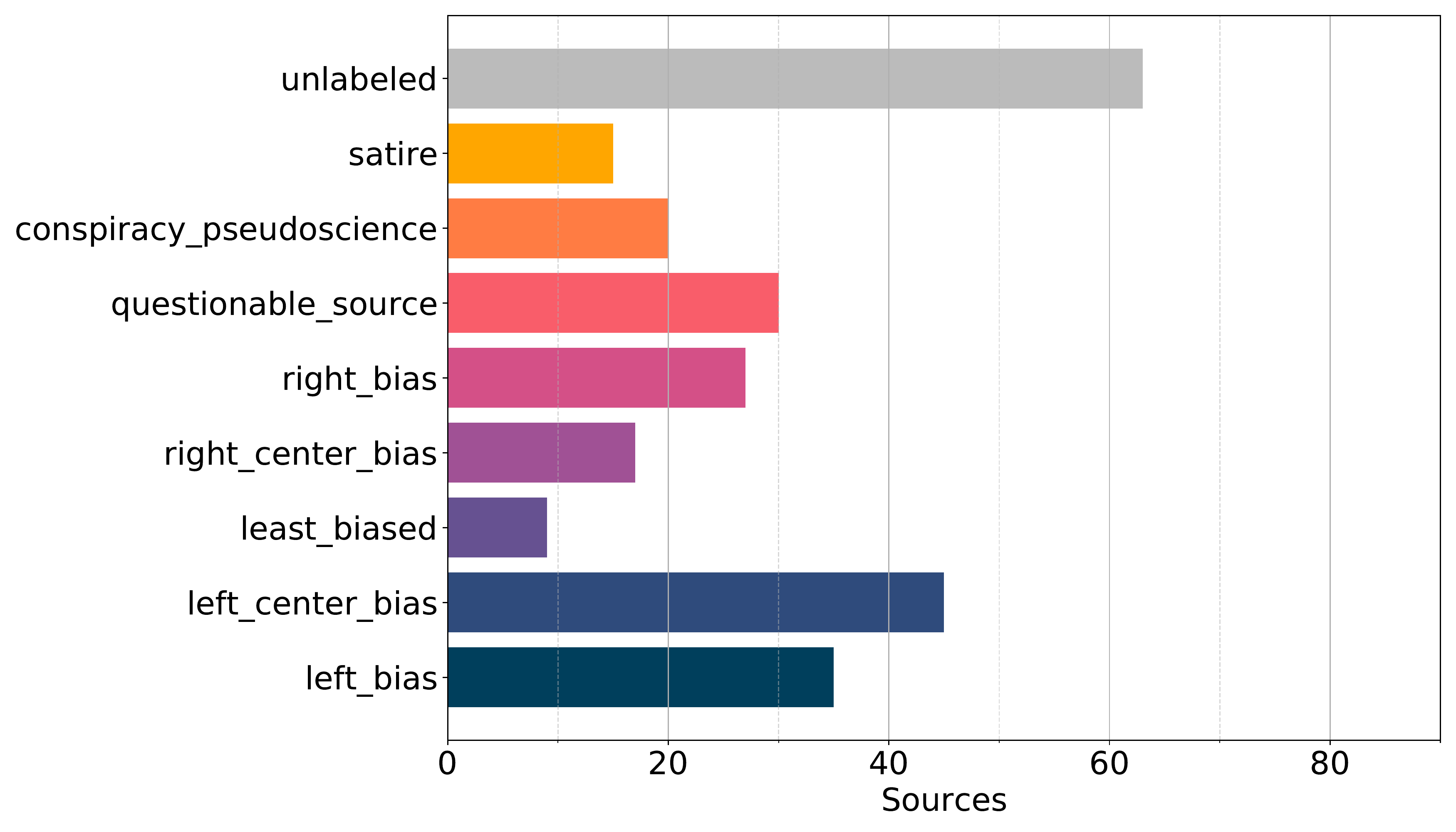}
        \caption{Sources per category}
    \end{subfigure}
    \caption{MBFC category distributions. In (a) we display the number of articles in each MBFC category, which include labels of political leaning and veracity. In (b) we display the number of sources in each MBFC category.}
    \label{fig:overtime_plot}
\end{figure*}

\begin{figure*}[ht!]

    \begin{subfigure}{\textwidth}
        \centering
        \includegraphics[width=0.8\textwidth]{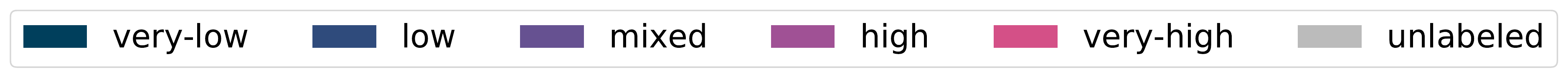}
    \end{subfigure}

    \begin{subfigure}{0.48\textwidth}
        \centering
        \includegraphics[width=\textwidth]{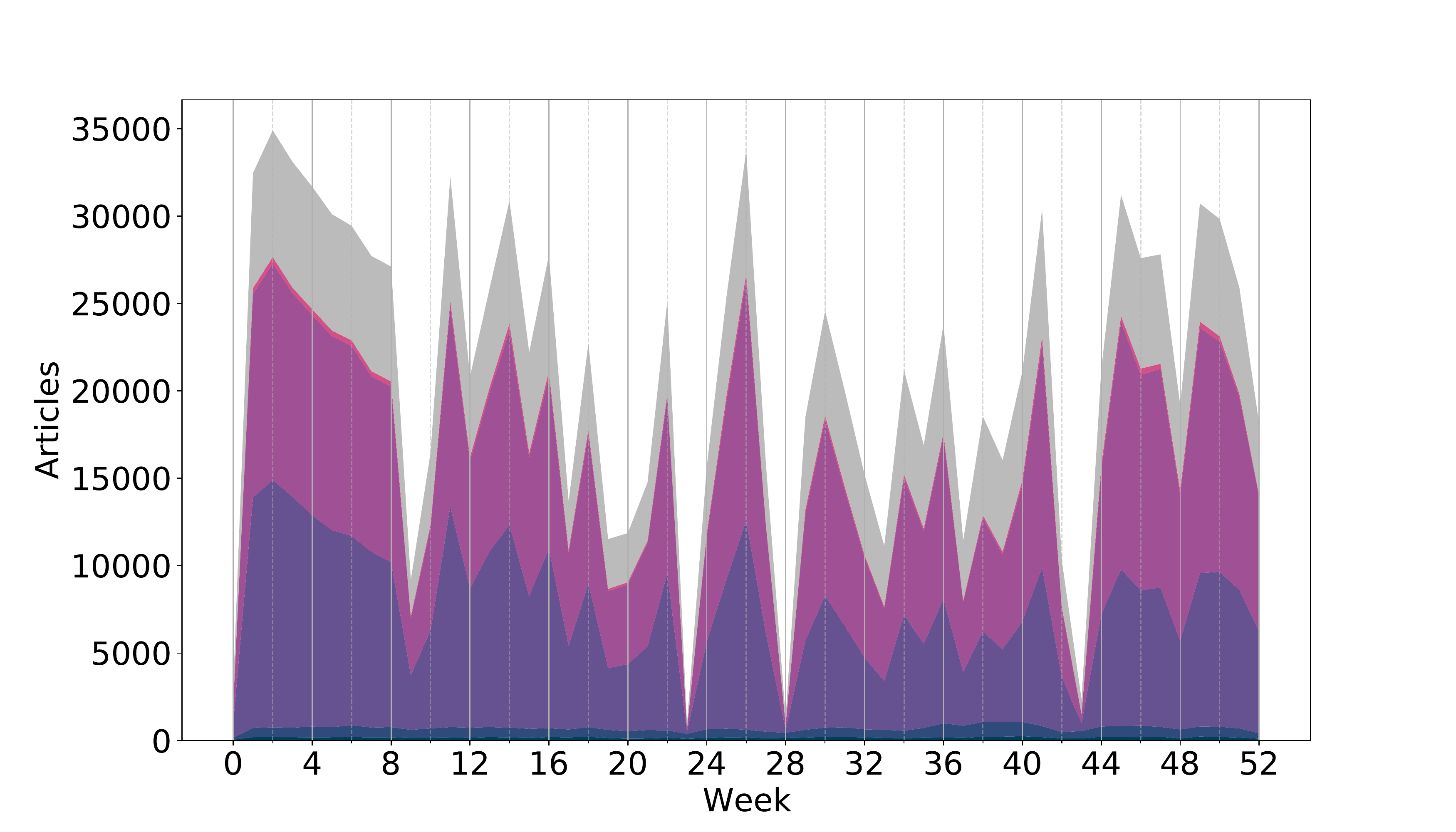}
        \caption{MBFC Factuality}
    \end{subfigure}
    \begin{subfigure}{0.48\textwidth}
        \centering
        \includegraphics[width=\textwidth]{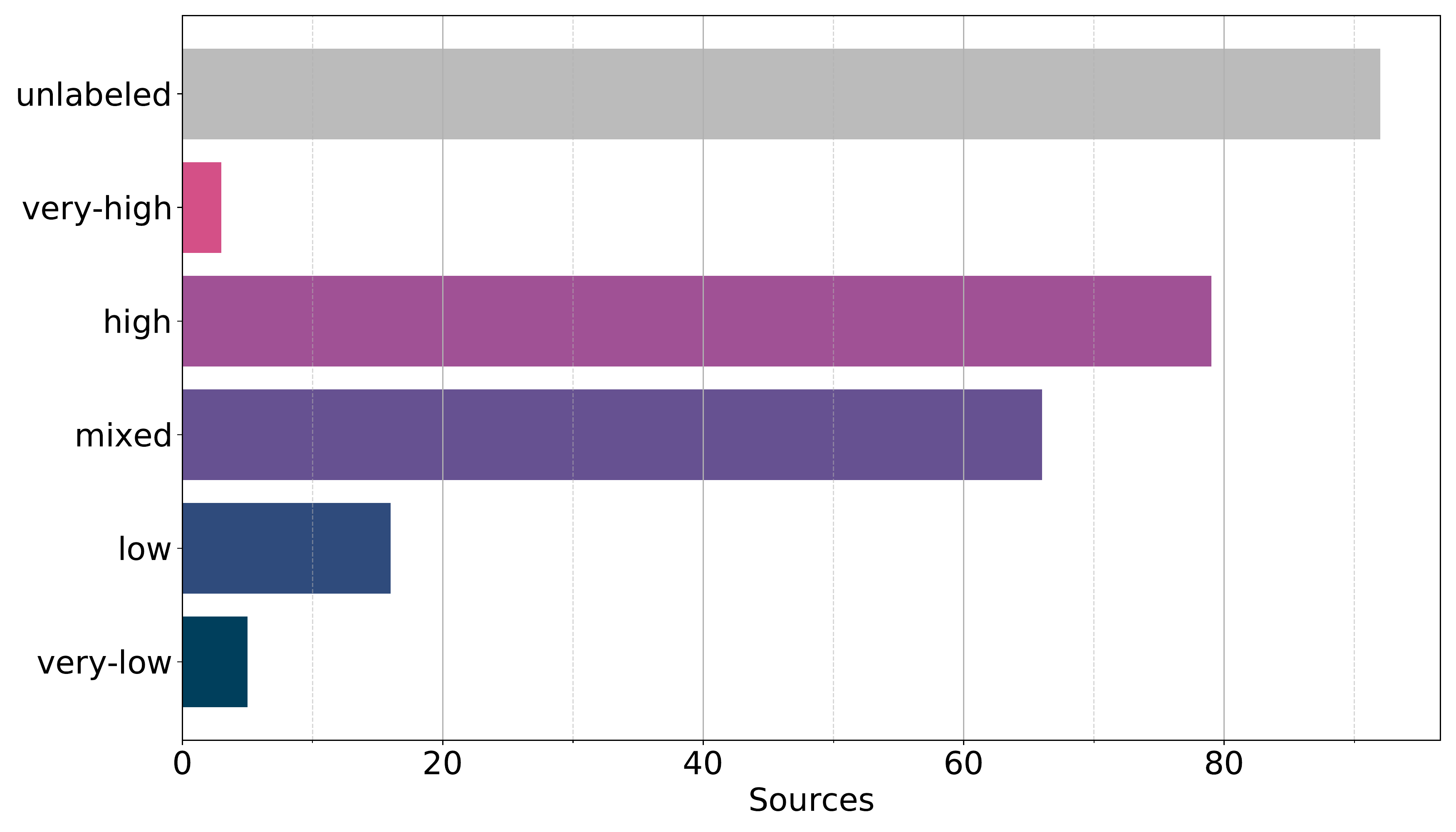}
        \caption{Sources per factuality level}
    \end{subfigure}
    \caption{MBFC facuality distributions. In (a) we display the number of sources in each in each MBFC factuality category, which is a range from very low to very high factuality. In (b) we display the number of sources in each category.}
    \label{fig:fact}
\end{figure*} 

\section{Whats New in NELA-GT-2019?}~\label{sec:whatsnew}
Other than being an updated time-frame, there are four primary differences between \texttt{NELA-GT-2019} and its previous version \texttt{NELA-GT-2018}.
\begin{enumerate}
    \item More data: We have added 66 more sources to our live collection, collecting approximately 400K more articles. Additionally, we have better stabilized our collection method, allowing us to collect more consistently over the year (see Figure~\ref{fig:overtime_plot}). Due to this increased stability, we have collected two more months of data than we did in 2018 (i.e. 10 months in \texttt{NELA-GT-2018} vs. 12 months in \texttt{NELA-GT-2019}). 
    \item Updated ground truth: We have updated our ground truth labels, particularly as it pertains to Media Bias/Fact Check (MBFC). Despite the large addition of new sources in the dataset, we have maintained a high density of source-level labels. Specifically, 79\% of the sources have at least 1 label from the 7 different assessment sites and 76\% of sources have a MBFC label. In \texttt{NELA-GT-2018}, we also had 79\% sources with at least 1 label, but with fewer sources in the collection. One major change in the labels provided is the removal of NewsGuard labels. Since the release of \texttt{NELA-GT-2018}, NewsGuard has moved to a paywall model and has change its terms of service accordingly. Hence, we have decided to remove their labels from the dataset. Also new to \texttt{NELA-GT-2019} is a 3-class aggregated label of source reliability, described in Section~\ref{sec:gtdata}. 
    \item New formats: We have released the data in two formats: (1) a SQLite database (2) a JSON dictionary per news source. In the past, we released the dataset in a SQLite database format and a plain text format. Due to the growth of the dataset, we have decided to move away from the plain text format to the JSON format. Details about the database schema and JSON dictionary format can be found in Section \ref{format}.
    \item Extraction code included: Also new in this years release are ready-to-go Python scripts for extracting the data from either the SQLite database format or the JSON format. 
\end{enumerate}

\begin{figure*}[ht!]

    \begin{subfigure}{\textwidth}
        \centering
        \includegraphics[width=0.6\textwidth]{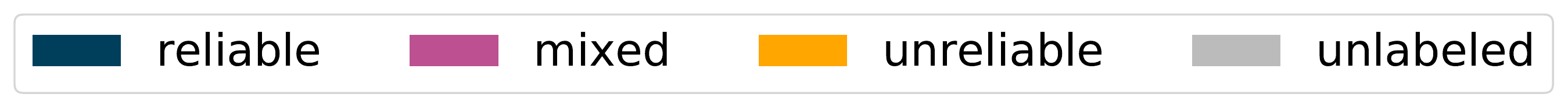}
    \end{subfigure}

    \begin{subfigure}{0.48\textwidth}
        \centering
        \includegraphics[width=\textwidth]{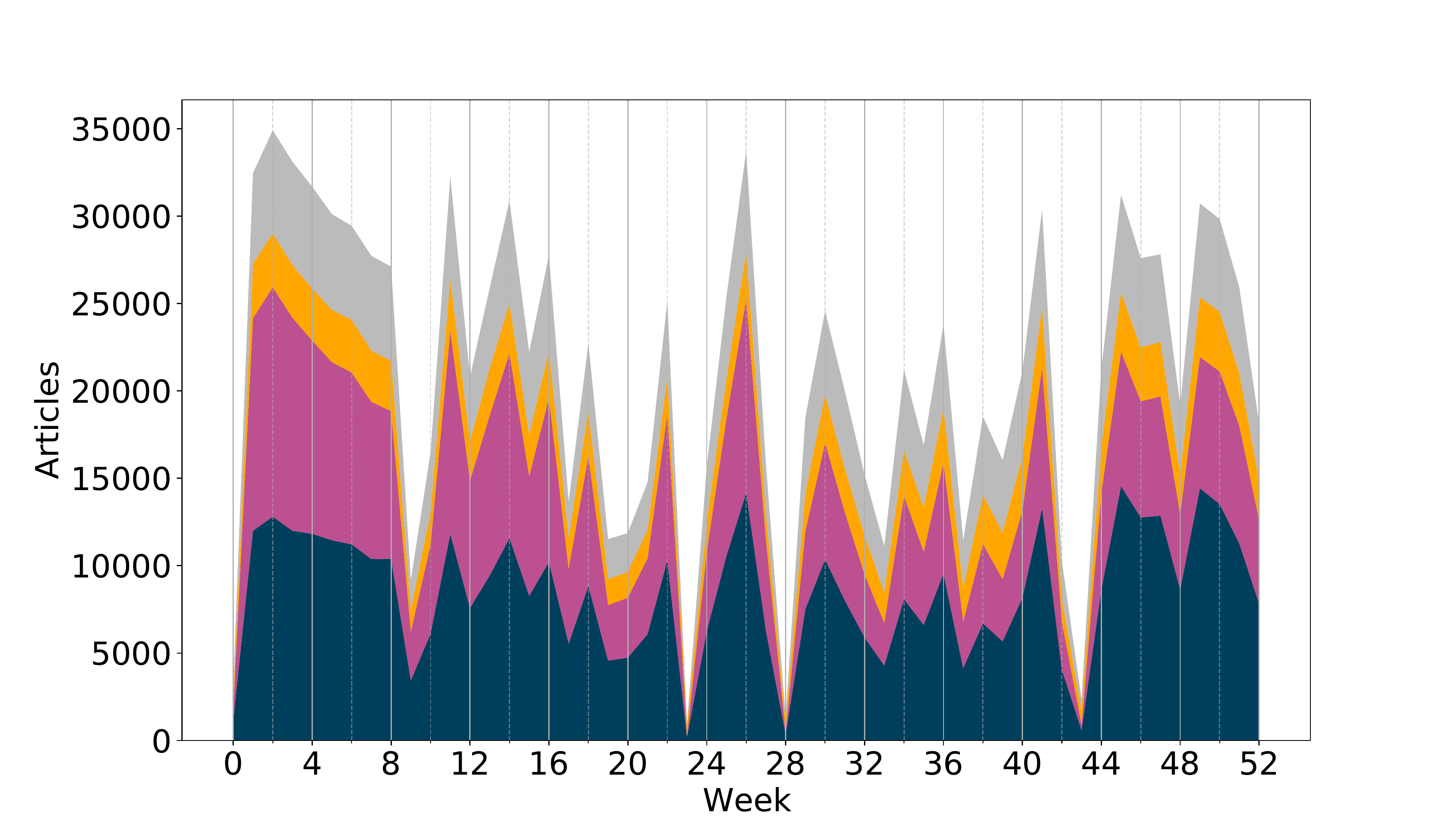}
        \caption{Articles per class}
    \end{subfigure}
    \begin{subfigure}{0.48\textwidth}
        \centering
        \includegraphics[width=\textwidth]{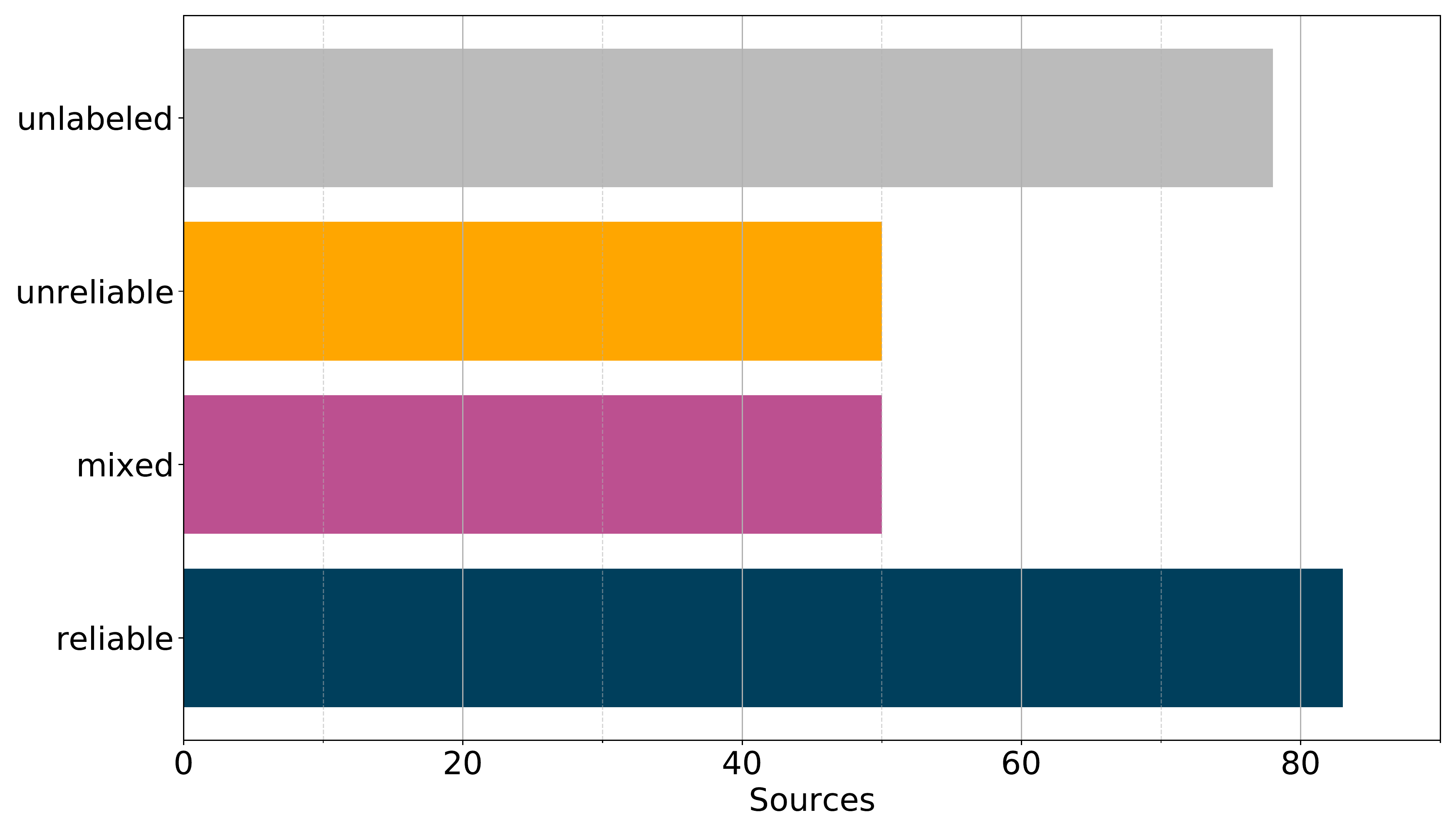}
        \caption{Sources per class}
    \end{subfigure}
    \caption{Distribution of the aggregated classes. In (a) we see the number of articles per class over time. In (b) we see the total number of sources in each class.}
    \label{fig:source_counts}
\end{figure*}  

\section{Data Collection}
The data collection process follows what was described in \cite{norregaard2019nela}. Specifically, we scraped the RSS feeds of
each source in our source collection list twice a day starting on 01/01/2019 using the Python libraries feedparser and goose. Our list of sources to collect was carried over from \cite{norregaard2019nela}, with an additional 66 sources added to this list. These additional sources mostly include conspiracy/pseudoscience news sites that have gained popularity over the past year. Just as in the 2018 version, these sources come from a variety of countries, but are all articles are in English.

\section{Format of Data}\label{format}
The dataset has been released in two formats: (1) a SQLite database (2) a JSON dictionary per news source. Details about the structure of each of these formats is below. We provide Python code to read both data formats at: \url{https://github.com/MELALab/nela-gt-2019}

\subsection{SQLite Database Format}

The SQLite 3 database format consists of a simple database with a single table called \texttt{newsdata}. This table contains the entire dataset, each row contains data about an article. Column \textbf{id} is set as primary key to avoid duplicated entries on the database. We normalized source names by converting them to lower case, and removing spaces, punctuation, and hyphens. For example, the source \emph{The New York Times} appears as \emph{thenewyorktimes},  Table \ref{tab:format} gives information about data columns.

\subsection{JSON Format}\label{json}

We also provide the dataset in JSON format. Specifically, each source has one JSON file containing the list of all of its articles. The fields follow the same structure of the database columns (Table \ref{tab:format}).

\begin{table*}[ht!]
    \centering
    \begin{tabular}{c|c|c}
         \textbf{Column} & \textbf{Type}  & \textbf{Description}\\ \toprule
         id & text (primary key) & article id \\
         date & text & publication date string in YYYY-MM-DD format \\
         source & text & name of the source from which the article was collected \\
         title & text & headline of the article \\
         content & text & body text of the article \\
         author & text & author of the article (if available) \\
         published & text & publication date time string as provided by source (inconsistent formatting) \\
         published\_utc & integer & publication time as unix time stamp \\
         collection\_utc & integer & collection time as unix time stamp
    \end{tabular}
    \caption{Structure of NELA-GT-2019 data. For the database format, column \textbf{id} is the primary key of table \texttt{newsdata}.}
    \label{tab:format}
\end{table*}

\section{Ground Truth Data}~\label{sec:gtdata}
Just as in \texttt{NELA-GT-2018}, we include multiple types of source-level labels. In \texttt{NELA-GT-2019}, we collect source-level labels from 7 different assessment sites:
\begin{enumerate}
    \item Media Bias/Fact Check (MBFC)
    \item Pew Research Center
    \item Wikipedia
    \item OpenSources
    \item AllSides
    \item BuzzFeed News
    \item Politifact
\end{enumerate}

As mentioned in Section~\ref{sec:whatsnew}, we removed NewsGaurd from our news assessment list (which was used in in the 2018 dataset) due to changes in their terms of service. Furthermore, some of these assessment sites no longer exist (OpenSources) or are not updated (Pew Research Center, BuzzFeed News), but labels are carried over from the 2018 dataset. The assessments that have been updated since 2018 are MBFC, AllSides, and Politifact. We refer the reader to the \texttt{NELA-GT-2018} paper for details on each assessment site~\cite{norregaard2019nela}.

Based on these 7 assessments, we also create aggregated 3-class label: unreliable, mixed, and reliable. This aggregated label is computed using two pieces of information from MBFC: the source type and the factual reporting score. Using source type, we label \texttt{unreliable} any source that has been flagged by MBFC as \emph{conspiracy} or \emph{pseudoscience}. Using the factual reporting score from MBFC, we label \texttt{unreliable} sources whose factual reporting is \emph{low} or \emph{very low}, \texttt{mixed} if the factual reporting is \emph{mixed}, and \texttt{reliable} if the factual reporting is \emph{high} or \emph{very high}. Thus, creating a three-class labeling of sources (0 - reliable, 1 - mixed, 2 - unreliable).

\subsection{Ground Truth Data Format}
Just as in the 2018 version of the dataset, we have ground truth data formatted as a CSV file, in which rows are sources and columns are ground truth types from the 7 different assessment sites. If a source has no labels, it will simply be the source name followed by an empty row. This CSV includes our aggregated label (called $aggregated\_label$). 

\section{Long-term Use Cases}
One of our goals with the continued release of the NELA datasets is to support long-term news research. For example, \texttt{NELA2017}~\cite{horne2018sampling}, \texttt{NELA-GT-2018}~\cite{norregaard2019nela}, and \texttt{NELA-GT-2019} can be combined to create a news article dataset covering over 2.5 years. With this 2.5 years of fairly consistent news data (or just the one year of data presented in this paper), there are several types of studies that can be performed:
\begin{itemize} 
    \item Concept drift in news veracity detection: Research on ``fake news'' detection has increased considerably in the past several years. However, much of this work has been on smaller, time-specific datasets. While this type of analysis is the first step in building news veracity models, understanding how stable the models performances are over time is crucial, particularly with automatic feature extraction methods which may overfit to time-specific features or topics. Using the dataset presented in this paper or the combination of the NELA datasets, this type of testing can be done. 
    \item Semi-supervised news veracity detection: While the NELA datasets have source-level labels for a majority of sources, there are many unlabeled sources in the dataset. Furthermore, while some sources are easily defined as reliable or unreliable, there are many mixed veracity sources. Can these unlabeled and mixed veracity sources be used in semi-supervised models? Semi-supervised and unsupervised models for news veracity have been explored, but remain under-explored in the literature.
    \item Disinformation producer tactics over time: While there has been substantial focus on ``fake news'' detection methods by researchers, there has been very little work on disinformation producer tactics. Of the studies that have focused on this, they have for, the most part, been focused on tactics during specific events or time-frames. Open questions in this area include: how do these tactics change over time? If tactics change over time, how can we account for those changes in our detection models? These types of questions can be answered using the NELA datasets. 
    \item Political narratives through events: Since the dataset (and combination of datasets) covers many major political events, studying how narratives change across each event and news source is possible. This type of analysis becomes important in understanding hyper-partisan news and its potential impacts on public opinion. 
\end{itemize}


\section{Conclusion}
In this short paper, we described the release of a 2019 labeled news article dataset for use in news veracity research. We provide a large dataset of news articles (1.2M articles), collected from 260 sources, over a one year (01/2018-12/2019). The articles are collected independent of social networks, thus are independent of specific community engagement. Due to this direct collection, the dataset approximately reflects the publishing patterns of each news source. In addition, we have included an array of source-level labels from 7 different assessment sites, each assessing the reliability or bias of a source. We have also included our own aggregated label based on these assessments. Lastly, we provide multiple data formats, code to extract the data, and use case examples to make working with the dataset easy. We hope that this dataset can continue to advance both computational and non-computational work in the field of news veracity. 

\bibliographystyle{aaai}
\bibliography{references}
\end{document}